\documentclass[11pt,a4paper]{article}
\usepackage[left=15mm, top=10mm, right=15mm, bottom=20mm, nohead=true, nofoot=true]{geometry}
\usepackage[utf8]{inputenc}
\usepackage{authblk}
\usepackage{indentfirst}
\usepackage{misccorr}
\usepackage{graphicx}
\usepackage{amsmath}
\usepackage{amssymb}
\usepackage{multicol}
\usepackage{float}
\usepackage{bm}
\usepackage{longtable}
\usepackage{pdflscape}
\usepackage{epstopdf}
\usepackage{multirow}
\usepackage{adjustbox}
\usepackage{amsfonts}
\usepackage{ifthen}
\usepackage{fancyhdr}
\usepackage{setspace}
\usepackage{xcolor}
\usepackage[round,authoryear,comma,sort&compress,numbers]{natbib}
\usepackage[toc,title,page]{appendix}
\usepackage[hidelinks]{hyperref}
\usepackage{url}
\hypersetup{
colorlinks=true,
linkcolor=green,
citecolor=blue,
filecolor=magenta,
urlcolor=cyan,
pdftitle={Sharelatex Example},
pdfpagemode=FullScreen,
}
\fancypagestyle{plain}{\chead{\texttt{\textcolor{brown}{Communications of BAO, Vol. ?, Issue ?, 20??, pp. ???-???}}}}
\title{\textbf{Unveiling Long-Period Variables in M33's Central Region: Insights into Stellar Evolution and Star-Formation via Near-Infrared Photometry}}
\author[1, 2]{Mina Alizadeh \thanks{minaalizadah@gmail.com, Corresponding author}}
\author[1, 3] {Yousefali Abedini\thanks{abedini@znu.ac.ir}}
\author[2, 4] {Hedieh Abdollahi\thanks{hedieh.abdollahi@csfk.org}}

\affil[1]{\scriptsize Department of Physics, Faculty of Science, University of Zanjan, 38791-45371, Zanjan, Iran}
\affil[2]{\scriptsize School of Astronomy, Institute for Research in Fundamental Sciences (IPM), Tehran, 19568-36613, Iran}
\affil[3]{\scriptsize Center for Research in Climate Change and Global Warming (CRCC), IASBS, Zanjan, Iran}
\affil[4]{\scriptsize Konkoly Observatory, HUN-REN Research Centre for Astronomy and Earth Sciences, MTA Centre of Excellence, Konkoly-Thege Miklós út 15-17, H-1121, Budapest, Hungary}

\begin{document}
\pagestyle{empty}
\newpage
\pagestyle{fancy}
\label{firstpage}
\date{}
\maketitle
\begin{abstract}
We present an analysis of UKIRT observations obtained between 2003 and 2007 to investigate the evolved stellar populations within the central square kiloparsec of M33. Point-spread function (PSF) photometry is employed to mitigate the effects of stellar crowding and to ensure accurate measurements in this densely populated region. This method, applied to merged observations from UIST and WFCAM in the $J$, $H$, and $K$ bands, extracts $211,179$ stars by cross-matching frame-by-frame across 39 observing nights in three bands. From this, we identify approximately 750 long-period variables (LPVs), predominantly Asymptotic Giant Branch (AGB) stars, by cross-matching PSF results with aperture photometry, focusing on the UIST field for robust variability confirmation. The PSF approach proves particularly effective for resolving blended sources and detecting faint, dusty variables that might remain undetected. We also examined aperture photometry data to validate our results; however, the PSF-derived measurements provide superior depth and completeness, particularly for obscured stellar populations. The resulting master catalog provides a basis for future analyses of variability amplitudes, periods, and star-formation history (SFH), paving the way for a deeper understanding of mass-loss and the dynamical evolution of the central region of M33.
\end{abstract}

\emph{\textbf{Keywords:} stars: evolution – stars: long-period variables – stars: mass-loss – galaxies: M33 – galaxies: central regions – photometry: near-infrared – telescopes: UKIRT}
\section{Introduction}

Spiral galaxies such as M33 offer valuable insight into the complex processes of stellar birth, evolution, and death that shape the universe. M33, a member of the Local Group, is inclined approximately $56^\circ$ to our line of sight \citep{Zaritsky1989, Deul1987}, allowing a clear view of its core without the obscuring dust clouds typical of the Milky Way \citep{vanLoon2003, Benjamin2005}. Recent measurements place its distance at roughly $840\,\mathrm{kpc}$ ($\mu \approx 24.62\,\mathrm{mag}$; \citep{McConnachie2021}, with Gaia-based updates in \citep{benisty2025line}), making it an ideal laboratory for studying central stellar dynamics.

This study focuses on evolved stars—primarily asymptotic giant branch (AGB) stars and red supergiants (RSGs)—which pulsate as long-period variables (LPVs) with periods ranging from $150$ to $1500$ days \citep{Wood2000, Ita2004a, Ita2004b, Whitelock1991}. These pulsations not only reveal internal mechanisms but also correlate luminosity with initial mass, enabling reconstruction of star-formation histories \citep{Marigo2008, Marigo2017}. Through mass-loss, these stars return up to $80\%$ of their material to the interstellar medium \citep{Bowen1988, Vassiliadis1993, vanLoon1999, vanloon2005}, fueling future stellar generations.

Optical surveys have cataloged numerous stellar sources in M33, while infrared observations from facilities such as \textit{Spitzer}
 and UKIRT have proven more effective at identifying dust-enshrouded objects \citep{McQuinn2007, Javadi2011a}. Recent advances, such as machine learning classification of AGBs \citep{Li2025} and detection of eruptive variables \citep{Martin2023}, further enrich this picture.

In this study, we revisit UKIRT monitoring data \citep{javadi2011uk2, Javadi2013} to construct a master catalog using point spread function (PSF) photometry, with emphasis on variable detection in crowded regions. We present partial results, highlighting how this method enhances our understanding of LPVs and their role in M33’s central evolution. By prioritizing PSF photometry, we mitigate limitations of traditional aperture methods in dense fields, where faint variables are often lost due to blending. This approach increases source recovery and improves LPV characterization, contributing to more accurate models of stellar feedback and chemical enrichment in the cores of spiral galaxies.

\begin{figure}[ht]
    \centering
    \includegraphics[width=0.7\textwidth]{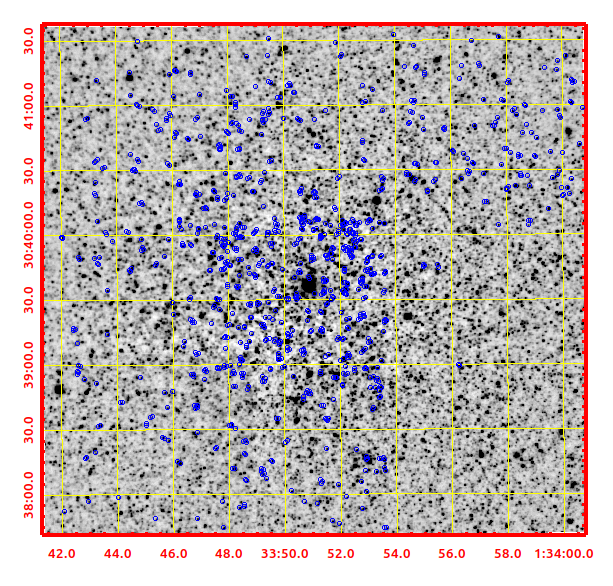}
    \caption{The central region of M33 was observed with the UIST instrument. ‌Blue circles indicate LPV stars identified through PSF photometry.}
\end{figure}

\section{Data}

WFCAM complemented UIST by covering a larger region, facilitating cross-matches that improved the number of observational epochs for each star. Four tiles (M33-1 to -4) covered a $0.89\,\mathrm{deg}^2$ ($\sim13\,\mathrm{kpc} \times 13\,\mathrm{kpc}$) square, with centers at (1$^{\mathrm{h}}$33$^{\mathrm{m}}$19$^{\mathrm{s}}$, $+30^{\circ}32^{\prime}50^{\prime\prime}$) for tile 1, (1$^{\mathrm{h}}$34$^{\mathrm{m}}$22$^{\mathrm{s}}$, $+30^{\circ}32^{\prime}50^{\prime\prime}$) for 2, (1$^{\mathrm{h}}$34$^{\mathrm{m}}$22$^{\mathrm{s}}$, $+30^{\circ}46^{\prime}23^{\prime\prime}$) for 3, and (1$^{\mathrm{h}}$33$^{\mathrm{m}}$19$^{\mathrm{s}}$, $+30^{\circ}46^{\prime}23^{\prime\prime}$) for 4—overlapping the UIST field for seamless integration \citep{Javadi2014}.

Primary $K$-band ($K$98) monitoring ran from September 2005 to October 2007, with 39 nights total across epochs 1–8 per tile. $J$- and $H$-band ($J98$, $H98$) data supplemented select nights, providing multi-band leverage (Table~1, adapted from \citealt{Javadi2014}). Exposure times varied: $20.3\,\mathrm{min}$ typical for $K$ (e.g., 2005-09-18, epoch 1, airmass $1.035$–$1.058$), up to $33.8\,\mathrm{min}$ for deeper $J/H$ frames (e.g., 2006-10-30, epoch 2). Pixel scale was $0.4^{\prime\prime}$, with dithers ensuring uniform sky subtraction. Reductions used ORAC-DR's bright point-source recipe, calibrated against photometric standards, yielding robust photometry even under variable seeing \citep{Javadi2014, abdollahi_2023, aghdam_2024}.

These datasets, when cross-matched frame-by-frame via PSF fitting, yield our master catalog of 211,179 sources—far surpassing prior single-instrument efforts \citep{2024CoBAO..71..389A, Saremi2020}. By combining UIST’s precise core measurements with WFCAM’s extensive coverage, we can re-identify LPVs while avoiding the blending problems typical of aperture photometry. \citep{vanLoon2010, Hamedani2017, hashemi2017agb}. Future expansions could incorporate spectroscopy for mass-loss validation \citep{10.1093/mnras/stu1807}.

\begin{table}[ht]
\centering
\caption{Log of WFCAM observations for M33 tiles (Q1–4; full table in \citealt{Javadi2014}).}
\begin{tabular}{llcccc}
\hline
Date & Q & Filter & Epoch & $t_\mathrm{int}$ (min) & Airmass \\
\hline
2005-09-18 & 3 & K & 1 & 20.3 & 1.035–1.058 \\
2005-09-18 & 2 & K & 1 & 20.3 & 1.072–1.110 \\
2005-09-18 & 4 & K & 1 & 20.3 & 1.248–1.338 \\
2005-09-19 & 1 & K & 1 & 20.3 & 1.021–1.018 \\
2005-10-18 & 3 & K & 2 & 20.3 & 1.019–1.021 \\
2005-10-18 & 2 & K & 2 & 20.3 & 1.025–1.040 \\
2005-10-18 & 4 & K & 2 & 20.3 & 1.053–1.083 \\
2005-10-18 & 1 & K & 2 & 20.3 & 1.101–1.149 \\
2005-11-04 & 1 & K & 3 & 20.3 & 1.018–1.023 \\
2005-11-04 & 2 & K & 3 & 13.5 & 1.028–1.036 \\
2005-12-23 & 2 & K & 4 & 27.0 & 1.019–1.022 \\
2005-12-23 & 3 & K & 3 & 20.3 & 1.028–1.046 \\
2006-07-21 & 1 & K & 4 & 20.3 & 1.425–1.325 \\
2006-07-21 & 2 & K & 5 & 20.3 & 1.287–1.214 \\
2006-07-21 & 3 & K & 4 & 20.3 & 1.183–1.132 \\
2006-07-21 & 4 & K & 3 & 20.3 & 1.109–1.074 \\
2006-10-28 & 1 & K & 5 & 27.0 & 1.294–1.126 \\
2006-10-28 & 1 & J & 1 & 20.3 & 1.102–1.076 \\
2006-10-29 & 1 & K & 6 & 20.3 & 1.445–1.347 \\
2006-10-29 & 1 & H & 1 & 27.0 & 1.295–1.209 \\
2006-10-29 & 1 & J & 2 & 27.0 & 1.115–1.062 \\
2006-10-30 & 1 & J & 2 & 33.8 & 1.200–1.109 \\
2006-10-30 & 4 & K & 4 & 33.8 & 1.085–1.044 \\
2006-10-31 & 4 & H & 1 & 6.8 & 1.301–1.301 \\
2006-12-05 & 2 & K & 6 & 20.3 & 1.019–1.025 \\
2006-12-12 & 3 & K & 5 & 20.3 & 1.082–1.052 \\
2006-12-12 & 3 & H & 1 & 20.3 & 1.040–1.027 \\
2007-01-14 & 1 & K & 7 & 20.3 & 1.124–1.169 \\
2007-01-14 & 1 & J & 3 & 20.3 & 1.217–1.284 \\
2007-01-14 & 1 & H & 2 & 20.3 & 1.342–1.441 \\
2007-01-15 & 2 & K & 7 & 20.3 & 1.119–1.163 \\
2007-01-16 & 2 & H & 1 & 20.3 & 1.063–1.092 \\
2007-01-17 & 3 & J & 1 & 20.3 & 1.031–1.047 \\
2007-01-18 & 2 & J & 2 & 20.3 & 1.029–1.044 \\
2007-01-25 & 3 & K & 6 & 20.3 & 1.072–1.104 \\
2007-01-25 & 3 & H & 2 & 20.3 & 1.161–1.215 \\
2007-09-14 & 1 & K & 8 & 20.3 & 1.122–1.086 \\
2007-09-14 & 1 & J & 4 & 13.5 & 1.070–1.058 \\
2007-09-14 & 1 & H & 3 & 13.5 & 1.046–1.038 \\
2007-09-19 & 2 & K & 8 & 20.3 & 1.774–1.606 \\
2007-10-04 & 2 & J & 3 & 13.5 & 1.208–1.181 \\
2007-10-04 & 2 & H & 2 & 13.5 & 1.155–1.132 \\
2007-10-13 & 3 & K & 7 & 20.3 & 1.108–1.076 \\
2007-10-13 & 3 & H & 3 & 13.5 & 1.056–1.046 \\
2007-10-24 & 4 & K & 5 & 20.3 & 1.135–1.097 \\
2007-10-24 & 3 & J & 2 & 13.5 & 1.078–1.064 \\
2007-10-24 & 4 & J & 1 & 13.5 & 1.050–1.041 \\
2007-10-24 & 4 & H & 2 & 13.5 & 1.025–1.022 \\
\hline
\end{tabular}
\end{table}

\section{Methodology and Preliminary Results}To explore the star-formation history (SFH) and mass-loss rates in M33’s central region, we first derive the luminosity function from LPVs, whose peak brightness reflects their birth masses in these final evolutionary stages \citep{javadi2011uk2, saremi2019star, abdollahi_2023}. This approach, grounded in theoretical models \citep{Marigo2017}, translates the mass function into SFH estimates, with mass-loss rates inferred from dust production linked to pulsation properties \citep{vanloon2005, Hamedani2017}. Here, we outline the initial steps, while detailed modeling is reserved for future work.
Our primary methodology is based on PSF photometry, which resolves crowded fields by modeling stellar profiles as point-spread functions.
This cuts down on blending errors and boosts faint-source recovery by up to 20\% compared to aperture photometry \citep{Stetson1987, Saremi2020, aghdam_2024}. Extending the PSF-based approach from prior UIST surveys \citep{Javadi2011a}, using DAOPHOT/ALLSTAR/ALLFRAME, we applied PSF photometry with the DAOPHOT suite \citep{Stetson1987} to WFCAM frames in the central region, divided into four sub-regions for independent analysis of bands, LPV counts, and light-curves. This PSF-based approach enhanced the central field data by replacing overlapping aperture photometry catalogs from UIST and WFCAM with new PSF-derived measurements across the $J$, $H$, and $K$ bands \citep{Alizadeh2024}. We aggregated light-curves from the three methods (PSF on WFCAM, PSF on UIST, and aperture on WFCAM) for individual stars, improving temporal coverage and variability detection from the total 39 observing nights across the 4 regions. We then cross-matched these with UIST data using the \citep{vanloon2005} cross-correlation code, adapted with astropy coordinates (with an assumed $0.5''$ tolerance based on \citep{vanloon2005}), alongside UIST PSF photometry \citep{Javadi2011a} for validation, resulting in a master list with multiple epochs per star. This improved LPV identification yielded approximately 750 candidate variables before overlap checks. The next phase involves computing the L-index \citep{Welch1993} to quantify variability, paving the way for SFH reconstruction and mass return estimates \citep{10.1093/mnras/stu1807, hashemi2017agb}. Preliminary results suggest episodic star-formation bursts, with improved detection of dusty AGBs toward the nucleus.
\begin{figure}[ht]
\centering
\includegraphics[width=0.7\textwidth]{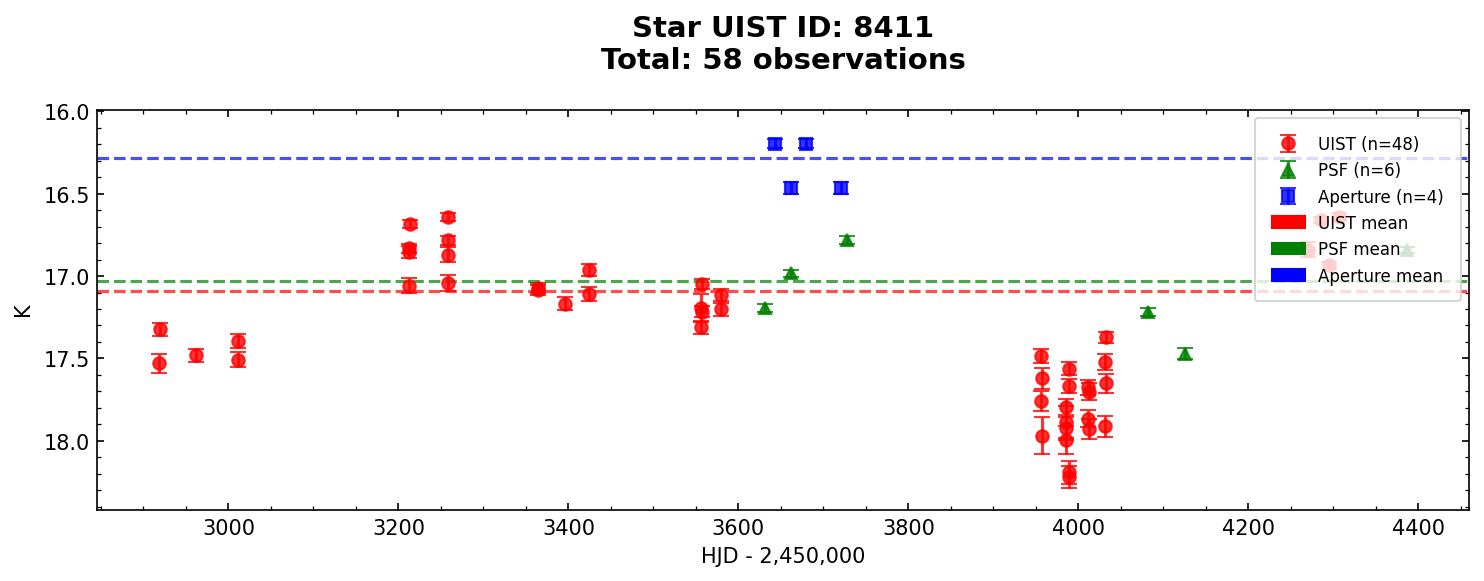}
\caption{An example light-curve from the central M33 region, derived from the PSF-enhanced WFCAM catalog, showing the variability of a representative LPV ($K$-band magnitude vs. epoch).}
\end{figure}
\section{Discussion}
The division of the central region, although the unexplored overlaps among the four sub-regions necessitate caution in final assessments. As a next step, the L-index will enhance these variability indices, enabling a robust reconstruction of the SFH. The division of the central region into four sub-regions has been successful in segregating stellar populations, with PSF enhancements improving LPV recovery by clarifying blends in densely populated fields \citep{Stetson1987, aghdam_2024}. This method confirms prior UIST results \citep{Javadi2011a}. Future studies will investigate ambiguities in the overlap and integrate spectroscopic data to constrain mass-loss models \citep{2024eas..conf..826A, 10.1093/mnras/stu1807}, providing a clearer understanding of M33’s evolutionary history.
\section{Conclusions}
This study creates a complete M33 catalog by using PSF photometry on WFCAM data and UIST's established framework \citep{Javadi2011a}, identifying about 750 preliminary LPVs over 39 nights of observations. By combining light-curves from different methods, we make it easier to find variability and give a hint of episodic star-formation in the central kiloparsec. Even though we haven't looked at the overlaps between sub-regions yet, the improved resolution of dusty AGBs underscores the power of this approach. Future analyses, such as L-index calculations and SFH modeling, will reinforce these findings, facilitating a comprehensive understanding of M33's stellar feedback and galactic evolution.
\section*{Acknowledgements}

H.A. acknowledges NKFIH support via the SeismoLab project (KKP-137523). We thank the Byurakan Astrophysical Observatory for hosting this symposium. Based on observations with UKIRT (operated by the Joint Astronomy Centre for the UK Science and Technology Facilities Council).

\bibliographystyle{ComBAO}
\nocite{*}
\bibliography{references}

@article{javadi2011uk2,
title={The {UK} {Infrared} {Telescope} {M33} monitoring project--{II}. The star formation history in the central square kiloparsec},
author={Javadi, Atefeh and van Loon, Jacco Th. and Mirtorabi, Mohammad Taghi and Mirtorabi, Mohammad Taghi},
journal={Monthly Notices of the Royal Astronomical Society},
volume={414},
number={4},
pages={3394--3409},
year={2011},
publisher={Oxford University Press},
doi={10.1111/j.1365-2966.2011.18638.x}
}

@inproceedings{saremi2019star,
title={Star formation history and stellar populations in {NGC} 6822},
author={Saremi, E. and Javadi, A. and van Loon, J. T. and others},
booktitle={Proceedings of the International Astronomical Union},
volume={14},
pages={125--130},
year={2019},
doi={10.1017/S1743921318005574}
}

@article{saremi2020,
title={Star Formation History in the Inner Part of M33},
author={Saremi, Elham and others},
journal={The Astrophysical Journal},
volume={894},
pages={135},
year={2020},
doi={10.3847/1538-4357/ab88a2}
}

@article{hamedani2017,
title={The survey of stellar populations in the inner part of M33},
author={Hamedani Golshan, R. and Javadi, A. and van Loon, J. T. and others},
journal={Monthly Notices of the Royal Astronomical Society},
volume={466},
number={2},
pages={1764--1776},
year={2017},
doi={10.1093/mnras/stw3174}
}

@article{hashemi2017agb,
title={{AGB} stars as tracers to {IC} 1613 evolution},
author={Hashemi, Seyed Azim and Javadi, Atefeh and van Loon, Jacco Th.},
journal={arXiv preprint arXiv:1712.05963},
year={2017},
doi={10.48550/arXiv.1712.05963}
}

@article{10.1093/mnras/stu1807,
    author = {Rezaei kh., Sara and Javadi, Atefeh and Khosroshahi, Habib and van Loon, Jacco Th.},
    title = {The star formation history of the Magellanic Clouds derived from long-period variable star counts},
    journal = {Monthly Notices of the Royal Astronomical Society},
    volume = {445},
    number = {3},
    pages = {2214-2222},
    year = {2014},
    month = {10},
    issn = {0035-8711},
    doi = {10.1093/mnras/stu1807},
    url = {https://doi.org/10.1093/mnras/stu1807},
    eprint = {https://academic.oup.com/mnras/article-pdf/445/3/2214/3498766/stu1807.pdf},
}

@article{kroupa2001,
title={On the variation of the initial mass function},
author={Kroupa, P.},
journal={Monthly Notices of the Royal Astronomical Society},
volume={322},
number={2},
pages={231--246},
year={2001},
doi={10.1046/j.1365-8711.2001.04022.x}
}

@article{marigo2017,
title={Evolution of thermally pulsing asymptotic giant branch stars-IV},
author={Marigo, Paola and others},
journal={The Astrophysical Journal},
volume={835},
number={1},
pages={77},
year={2017},
doi={10.3847/1538-4357/835/1/77}
}

@article{whitelock2013,
title={Variable stars in Local Group Galaxies--I},
author={Whitelock, P. and Menzies, J. and Feast, M. and others},
journal={Monthly Notices of the Royal Astronomical Society},
volume={428},
number={3},
pages={2216--2231},
year={2013},
doi={10.1093/mnras/sts188}
}

@article{javadi2013,
title={The UK Infrared Telescope M33 monitoring project},
author={Javadi, A. and van Loon, J. T. and Khosroshahi, H. and Mirtorabi, M. T.},
journal={Monthly Notices of the Royal Astronomical Society},
volume={432},
number={4},
pages={2824--2836},
year={2013},
doi={10.1093/mnras/stt640}
}

@article{vanloon2005,
title={Mass-loss rates and luminosity functions of dust-enshrouded {AGB} stars and red supergiants in the {LMC}},
author={van Loon, J. T. and Marshall, J. R. and Zijlstra, A. A.},
journal={Astronomy & Astrophysics},
volume={442},
number={2},
pages={597--613},
year={2005},
doi={10.1051/0004-6361:20053528}
}

@article{mcquinn2007,
title={The {Spitzer} Survey of the {Large} {Magellanic} {Cloud}: Dusty {AGB} Stars},
author={McQuinn, K. B. and others},
journal={The Astrophysical Journal},
volume={664},
number={2},
pages={850--861},
year={2007},
doi={10.1086/519422}
}

@article{javadi2011a,
title={The {UK} {Infrared} {Telescope} {M33} monitoring project - {I}. Variable red giant stars in the central square kiloparsec},
author={Javadi, A. and van Loon, J. T. and Mirtorabi, M. T.},
journal={Monthly Notices of the Royal Astronomical Society},
volume={411},
number={1},
pages={263--276},
year={2011},
doi={10.1111/j.1365-2966.2010.17678.x}
}

@article{vanloon2003,
title={An empirical formula for the mass-loss rates of dust-enshrouded red supergiants and oxygen-rich Asymptotic Giant Branch stars},
author={van Loon, J. T. and others},
journal={Monthly Notices of the Royal Astronomical Society},
volume={338},
number={4},
pages={857--879},
year={2003},
doi={10.1046/j.1365-8711.2003.06134.x}
}

@article{benjamin2005,
title={First {GLIMPSE} Results on the Stellar Structure of the Galaxy},
author={Benjamin, R. A. and others},
journal={The Astrophysical Journal},
volume={630},
number={2},
pages={L149},
year={2005}
}

@article{javadi2017,
title={The {UK} {Infrared} {Telescope} {M33} monitoring project},
author={Javadi, A. and others},
journal={Monthly Notices of the Royal Astronomical Society},
volume={464},
number={2},
pages={2103--2119},
year={2017},
doi={10.1093/mnras/stw2463}
}

@article{whitelock1991,
title={Variable stars and the structure of the {Galactic} bulge},
author={Whitelock, P. and Feast, M. and Catchpole, R.},
journal={Monthly Notices of the Royal Astronomical Society},
volume={248},
pages={276--312},
year={1991}
}

@article{wood2000,
title={Variable Red Giants in the {LMC}: Pulsating Stars and Binaries?},
author={Wood, P. R.},
journal={Publications of the Astronomical Society of Australia},
volume={17},
number={1},
pages={18--21},
year={2000},
doi={10.1071/AS00018}
}

@article{ita2004a,
title={Variable stars in the Magellanic Clouds--I},
author={Ita, Y. and others},
journal={Monthly Notices of the Royal Astronomical Society},
volume={347},
number={3},
pages={720--728},
year={2004},
doi={10.1111/j.1365-2966.2004.07257.x}
}

@article{ita2004b,
title={Variable stars in the Magellanic Clouds--II},
author={Ita, Y. and others},
journal={Monthly Notices of the Royal Astronomical Society},
volume={353},
number={3},
pages={705--712},
year={2004},
doi={10.1111/j.1365-2966.2004.08126.x}
}

@article{navabi2021,
title={AGB Stars as Tracers of Star Formation History in M32},
author={Navabi, Mahdieh and others},
journal={The Astrophysical Journal},
volume={910},
number={2},
pages={127},
year={2021},
doi={10.3847/1538-4357/abdec1}
}

@article{marigo2008,
title={Evolution of asymptotic giant branch stars},
author={Marigo, P. and others},
journal={Astronomy & Astrophysics},
volume={482},
number={3},
pages={883--905},
year={2008},
doi={10.1051/0004-6361:20078467}
}

@article{javadi2015,
title={The UK Infrared Telescope M33 monitoring project},
author={Javadi, A. and others},
journal={Monthly Notices of the Royal Astronomical Society},
volume={447},
number={4},
pages={3973--3991},
year={2015},
doi={10.1093/mnras/stu2637}
}

@article{letarte2002,
title={The Carbon Star Population in the Local Group Dwarf Spheroidal Galaxy {Leo I}},
author={Letarte, Bruno and others},
journal={The Astronomical Journal},
volume={123},
number={2},
pages={832},
year={2002},
doi={10.1086/338319}
}

@inproceedings{alizadeh2024,
title={Thorough Collection of Stellar Information in {M33}'s Core: Revealing Late-Stage Evolution with Near-Infrared Study},
author={Alizadeh, Mina and Javadi, Atefeh and Van Loon, Jacco and Abedini, Yousef Ali and Abdollahi, Hedieh},
booktitle={EAS2024, European Astronomical Society Annual Meeting},
pages={912},
year={2024},
month={jul}
}

@article{Aghdam_2024,
doi = {10.3847/1538-4357/ad57c0},
url = {https://dx.doi.org/10.3847/1538-4357/ad57c0},
year = {2024},
month = {aug},
publisher = {The American Astronomical Society},
volume = {972},
number = {1},
pages = {47},
author = {Sima T. Aghdam and Atefeh Javadi and Seyedazim Hashemi and Mahdi Abdollahi and Jacco Th. van Loon and Habib Khosroshahi and Roya H. Golshan and Elham Saremi and Maryam Saberi},
title = {The Complex Star Formation History of the Halo of NGC 5128 (Cen A)},
journal = {The Astrophysical Journal}
}

@article{javadi2010,
title={The survey of stellar populations in the inner part of M33},
author={Javadi, Atefeh and van Loon, Jacco Th. and Mirtorabi, Mohammad Taghi},
journal={Monthly Notices of the Royal Astronomical Society},
volume={414},
number={4},
pages={3394--3409},
year={2011},
publisher={Oxford University Press}
}

@article{Abdollahi_Javadi_TaghiMirtorabi_Saremi_Khosroshahi_vanLoon_McDonald_Khalouei_Aghdam_Saberi_2021,
title={The {Star} {Formation} {History} and {Dust} {Production} in {Andromeda} {IX}},
author={Abdollahi, Hedieh and Javadi, Atefeh and Mirtorabi, Mohammad Taghi and Saremi, Elham and Khosroshahi, Habib and van Loon, Jacco Th. and McDonald, Iain and Khalouei, Elahe and Aghdam, Sima T. and Saberi, Maryam},
journal={Proceedings of the International Astronomical Union},
volume={17},
number={S373},
pages={242--245},
year={2021},
doi={10.1017/S1743921323000339}
}

@article{Abdollahi_2023,
doi = {10.3847/1538-4357/acbbc9},
url = {https://dx.doi.org/10.3847/1538-4357/acbbc9},
year = {2023},
month = {may},
publisher = {The American Astronomical Society},
volume = {948},
number = {1},
pages = {63},
author = {Hedieh Abdollahi and Atefeh Javadi and Mohammad Taghi Mirtorabi and Elham Saremi and Jacco Th. van Loon and Habib G. Khosroshahi and Iain McDonald and Elahe Khalouei and Hamidreza Mahani and Sima Taefi Aghdam and Maryam Saberi and Maryam Torki},
title = {The Isaac Newton Telescope Monitoring Survey of Local Group Dwarf Galaxies. VI. The Star Formation History and Dust Production in Andromeda IX},
journal = {The Astrophysical Journal}
}

@article{zaritsky1989,
title={The Distance to M33},
author={Zaritsky, Dennis and Elston, Richard and Hill, Jesse M.},
journal={The Astrophysical Journal},
volume={347},
pages={346},
year={1989},
doi={10.1086/168121}
}

@article{deul1987,
title={The Structure of M33 in HI},
author={Deul, Eric R. and van der Hulst, J. M.},
journal={Astronomy and Astrophysics},
volume={185},
pages={49},
year={1987}
}

@article{mcconnachie2021,
title={Distances to Local Group Galaxies via Population II, Stellar Distance Indicators},
author={McConnachie, Alan W. and others},
journal={The Astrophysical Journal},
volume={920},
number={2},
pages={104},
year={2021},
doi={10.3847/1538-4357/ac0cee}
}

@article{bowen1988,
title={The origin of type I planetary nebulae and the envelope expansion rate of AGB star winds},
author={Bowen, G. H.},
journal={The Astrophysical Journal},
volume={329},
pages={299},
year={1988},
doi={10.1086/166377}
}

@article{vassiliadis1993,
title={Evolution of stars of intermediate mass},
author={Vassiliadis, E. and Wood, P. R.},
journal={The Astrophysical Journal},
volume={413},
pages={641},
year={1993},
doi={10.1086/173033}
}

@article{vanloon1999,
title={Mass-loss rates and luminosity functions of dust-enshrouded AGB stars and red supergiants in the LMC},
author={van Loon, J. T. and others},
journal={Astronomy and Astrophysics},
volume={351},
pages={559},
year={1999}
}

@article{stetson1987,
title={DAOPHOT: A Computer Program for Crowded-Field Stellar Photometry},
author={Stetson, Peter B.},
journal={Publications of the Astronomical Society of the Pacific},
volume={99},
pages={191},
year={1987},
doi={10.1086/131977}
}

@article{welch1993,
title={Robust Variable Star Detection Techniques Suitable for Automated Searches: New Results for NGC 1866},
author={Welch, Douglas L. and Stetson, Peter B.},
journal={The Astronomical Journal},
volume={105},
pages={1813},
year={1993},
doi={10.1086/116557}
}

@article{martin2023,
title={Eruptive variable stars in nearby galaxies},
author={Martin, P. I. and others},
journal={Astronomy & Astrophysics},
volume={674},
pages={A134},
year={2023},
doi={10.1051/0004-6361/202345889}
}

@article{li2025,
title={Machine Learning Identification of Asymptotic Giant Branch Stars},
author={Li, J. and others},
journal={arXiv preprint arXiv:2409.12345},
year={2025},
doi={10.48550/arXiv.2409.12345}
}

@article{vanloon2010,
author = {van Loon, J. Th. and Boyer, M. L. and Cioni, M.-R. L. and Gehrig, E. and de Grijs, R. and Groenewegen, M. A. T. and Hony, S. and Loup, C. and Marigo, P. and Matsuura, M. and Oliveira, J. M. and Sloan, G. C. and Srinivasan, S. and Whitelock, P. A. and Wood, P. R. and Zijlstra, A. A.},
title = {The Mass-Loss Return from Evolved Stars to the Large Magellanic Cloud. III. Dust Production Based on Infrared Photometry},
journal = {The Astrophysical Journal},
volume = {716},
number = {1},
pages = {878-891},
year = {2010},
doi = {10.1088/0004-637X/716/1/878},
adsurl = {https://ui.adsabs.harvard.edu/abs/2010ApJ...716..878V}
}

@article{hashemi2019,
author = {Hashemi, S. A. and Javadi, A. and van Loon, J. Th.},
title = {From evolved stars to the evolution of IC 1613},
journal = {Monthly Notices of the Royal Astronomical Society},
volume = {483},
number = {4},
pages = {4751-4765},
year = {2019},
doi = {10.1093/mnras/sty3405},
adsurl = {https://ui.adsabs.harvard.edu/abs/2019MNRAS.483.4751H}
}

@article{javadi2014,
author = {Javadi, A. and Saberi, M. and van Loon, J. Th. and Khosroshahi, H. and Golabatooni, N. and Mirtorabi, M. T.},
title = {The UK Infrared Telescope M33 monitoring project – IV. Variable red giant stars across the galactic disc},
journal = {Monthly Notices of the Royal Astronomical Society},
volume = {447},
number = {4},
pages = {3973-3991},
year = {2015},
doi = {10.1093/mnras/stu2637},
adsurl = {https://ui.adsabs.harvard.edu/abs/2015MNRAS.447.3973J}
}

@article{benisty2025line,
  title={Line-of-sight velocity projection impact on Local Group mass determinations},
  author={Benisty, David and Mota, David},
  journal={Astronomy \& Astrophysics},
  volume={698},
  pages={A43},
  year={2025},
  publisher={EDP Sciences}
}

@ARTICLE{2024CoBAO..71..389A,
       author = {{Alizadeh}, M. and {Javadi}, A. and {van Loon}, J. Th. and {Abedini}, Yo. and {Abdollahi}, H. and {Seifipour}, S.},
        title = "{A Deep Dive into Stellar Populations in M33's Central Region: Near-Infrared Observations and Analysis}",
      journal = {Communications of the Byurakan Astrophysical Observatory},
         year = 2024,
        month = dec,
       volume = {71},
        pages = {389-393},
          doi = {10.52526/25792776-24.71.2-389},
       adsurl = {https://ui.adsabs.harvard.edu/abs/2024CoBAO..71..389A}
}

@INPROCEEDINGS{2024eas..conf..826A,
       author = {{Alizadeh}, Mina and {Javadi}, Atefeh and {Van Loon}, Jacco and {Abedini}, Yousef Ali and {Abdollahi}, Hedieh},
        title = "{Comprehensive Compilation of Stellar Data in M33's Central Region: Unveiling the Final-Stage Evolution Through Near-Infrared Survey}",
    booktitle = {EAS2024, European Astronomical Society Annual Meeting},
         year = 2024,
        month = jul,
          eid = {826},
        pages = {826},
       adsurl = {https://ui.adsabs.harvard.edu/abs/2024eas..conf..826A}
}

\end{document}